\documentclass{emulateapj}

\shorttitle{Velocity Characteristics of Evaporated Plasma Using Hinode/EIS}

\shortauthors{Milligan \& Dennis}

\begin{document}

\title{Velocity Characteristics of Evaporated Plasma Using Hinode/EIS}

\notetoeditor{the contact email is ryan.o.milligan@nasa.gov and is the only one which should
appear on the journal version}

\author{Ryan O. Milligan\altaffilmark{1} \& Brian R. Dennis\altaffilmark{1}}

\altaffiltext{1} {Solar Physics Laboratory (Code 671), Heliophysics Science Division, NASA Goddard Space Flight Center, Greenbelt, MD 20771, U.S.A.}

\begin{abstract}
This paper presents a detailed study of chromospheric evaporation using the EUV Imaging Spectrometer (EIS) onboard Hinode in conjunction with HXR observations from RHESSI. The advanced capabilities of EIS were used to measure Doppler shifts in 15 emission lines covering the temperature range $T$=0.05--16~MK during the impulsive phase of a C-class flare on 2007 December 14. Blueshifts indicative of the evaporated material were observed in six emission lines from \ion{Fe}{14--XXIV} (2--16~MK). Upflow velocity ($v_{up}$) was found to scale with temperature as $v_{up}$ (km~s$^{-1}) \approx$ 8$-$18 $T$ (MK). Although the hottest emission lines, \ion{Fe}{23} and \ion{Fe}{24}, exhibited upflows of $>$200~km~s$^{-1}$, their line profiles were found to be dominated by a stationary component in contrast to the predictions of the standard flare model. Emission from \ion{O}{6}--\ion{Fe}{13} lines (0.5--1.5~MK) was found to be redshifted by $v_{down}$ (km~s$^{-1}) \approx$ 60$-$17 $T$ (MK) and was interpreted as the downward-moving `plug' characteristic of explosive evaporation. These downflows occur at temperatures significantly higher than previously expected. Both upflows and downflows were spatially and temporally correlated with HXR emission observed by RHESSI that provided the properties of the electron beam deemed to be the driver of the evaporation. The energy flux of the electron beam was found to be $\gtrsim$5$\times$10$^{10}$~ergs~cm$^{-2}$~s$^{-1}$ consistent with the value required to drive explosive chromospheric evaporation from hydrodynamic simulations.
\end{abstract}


\section{INTRODUCTION}
\label{intro}

Chromospheric evaporation is largely accepted to be the process by which solar flares produce their high-temperature, high-density plasma. The standard flare model states that electrons are accelerated at or near a magnetic reconnection site in the corona and then propagate along newly reconnected magnetic field lines towards the chromosphere. Here they are decelerated by the increasingly dense atmosphere and typically lose their energy by one of two mechanisms. An encounter with a proton or ion will result in the emission of a Hard X-Ray (HXR) photon through the bremsstrahlung process. Coulomb collisions with ambient electrons, on the other hand, result in an overall heating, and by consequence, expansion of the chromospheric material. The ratio of bremsstrahlung to collisional losses is of the order of 1:10$^{5}$.

The rate of expansion (or evaporation velocity) has traditionally been detected through Doppler measurements of Extreme Ultra-Violet (EUV) and Soft X-Ray (SXR) emission lines. \cite{anto83}, \cite{zarr88}, \cite{canf87}, \cite{acto82}, and \cite{dosc05} consistently measured blueshifts of 300--400~km~s$^{-1}$ in the \ion{Ca}{19} line (3.1--3.2\AA, 25~MK) using the Bent and Bragg Crystal Spectrometers (BCS) onboard SMM \citep{acto81} and Yohkon \citep{culh91}, respectively. In each case the \ion{Ca}{19} line profile was generally dominated by a component with a centroid position comparable to that of the rest wavelength while the upflows were indicated by a blue-wing enhancement. As BCS spectra were spatially integrated over the entire solar disk, \cite{dosc05} concluded that the stationary emission emanated from the top of the flaring loop while the blueshifted emission came from the footpoints. Similar studies using data from the Coronal Diagnostic Spectrometer (CDS; \citealt{harr95}) on SOHO revealed upflow velocities of 150--300~km~s$^{-1}$ in the \ion{Fe}{19} line (592.23\AA, 8~MK; \citealt{czay01},\citealt{teri03}, \citealt{bros04}, \citealt{mill06a,mill06b}). In these instances, the rastering capability of CDS showed that the emission known to come from loop footpoints was often completely blueshifted, in agreement with the interpretation of \cite{dosc05}. \ion{Ca}{19} and \ion{Fe}{19} are formed at distinctly different temperatures and consistently exhibit different upflow velocities. This suggests a direct relationship between the temperature and the expansion velocity of evaporated material. Indeed, most evaporation models predict this to be the case \citep{fish85c,fish85b,fish85a,mari89,allr05}.

\begin{figure*}[!ht]
\begin{center}
\includegraphics[height= 16cm,angle=90]{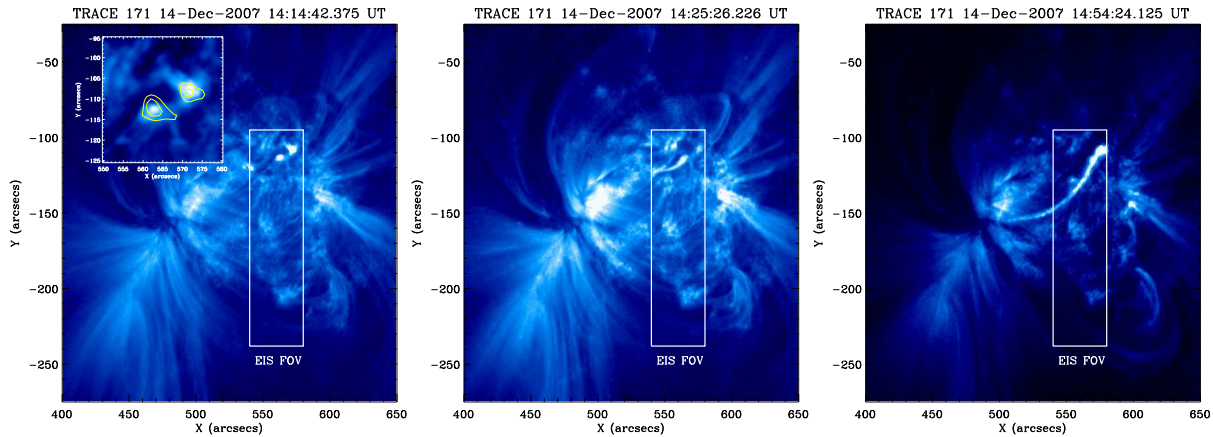}
\caption{Images of NOAA AR 10978 taken in the TRACE 171~\AA~passband during the different stages of the event on 2007 December 14. The first panel at 14:14:42~UT was taken during the impulsive phase. The inset in the top left corner shows a zoomed-in portion of the image containing the two HXR footpoints of the flare under investigation. The contours overlaid in yellow are the 60\% and 80\% levels of the 20--25~keV emission as observed by RHESSI from 14:14:28--14:15:00~UT. The second and third panels taken at 14:25:26 and 14:54:24~UT, respectively, show the flare at two different stages during the decay phase. Overlaid on each panel is the rectangular field of view of the EIS raster.}
\label{trace_hsi_eis_fov}
\end{center}
\end{figure*}

The velocity at which the evaporated material rises is dependent on the total energy flux of accelerated electrons (in ergs~cm$^{-2}$~s$^{-1}$) that reach the footpoints. Combining observations from CDS and the Reuven Ramaty High Energy Solar Spectroscopic Imager (RHESSI; \citealt{lin02}), \citet{mill06b} showed that for two flares with electron beam fluxes that differed by an order of magnitude, upflow velocities in the \ion{Fe}{19} line differed by a factor of two. If the incoming flux of electrons is above a given threshold (i.e. the heating rate is greater than the hydrodynamic expansion time scale), then the pressure of the hot, rising material becomes greater than that of the ambient chromosphere resulting in a `recoil' of the denser layers below, required to conserve momentum \citep{canf87}. \cite{fish85a} termed this as `explosive' (as opposed to `gentle') evaporation and deduced that the electron flux required was $\geq$3$\times$10$^{10}$~ergs~cm$^{-2}$~s$^{-1}$. Observational signatures of such downward motions (known as chromospheric condensation) are redshifts in chromospheric and transition region lines, formed below the electron deposition site (e.g. \citealt{mill06a,delz06}). These flows are distinctly different from similar downflows observed in coronal lines during the decay phase of the flare when the evaporated plasma eventually cools and precipitates back down the loop \citep{bros03,raft09}.

With the launch of Hinode \citep{kosu07} in September 2006, it is now possible to investigate the process of chromospheric evaporation by observing several high-temperature lines simultaneously with the EUV Imaging Spectrometer (EIS; \citealt{culh07}). EIS provides the capability to determine the spatial location of both the shifted and unshifted components at high spatial, spectral, and temporal resolution. Coincident HXR observations from RHESSI establish the location and the parameters of the driving electron beam that are needed to test the consistency with the model predictions.

This study focuses on diagnosing the evaporating plasma during the impulsive phase of a C-class flare. A general description of the flare is given in \S~\ref{obs}. The EIS observations, emission line fitting procedures, and velocity analysis are described in \S~\ref{eis_obs}. \S~\ref{rhessi} describes the RHESSI data analysis procedures while \S~\ref{conc} provides interpretations of the results in the context of the standard flare model. \S~\ref{future} presents some suggestions for future work.

\section{2007 December 14}
\label{obs}

\begin{figure}[!b]
\begin{center}
\includegraphics[height=8.5cm,angle=90]{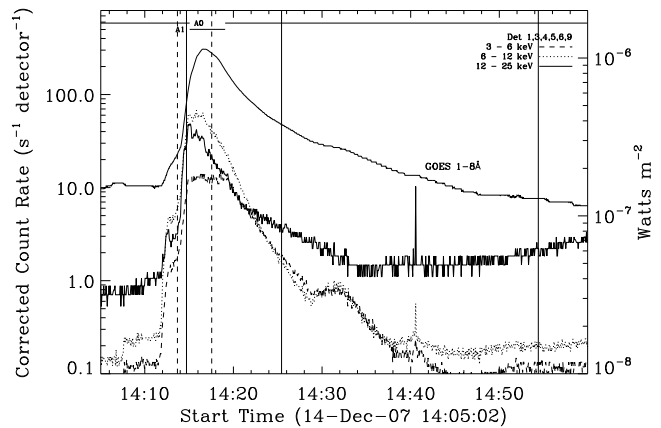}
\caption{RHESSI lightcurves of the event in the 3--6, 6--12, and 12--25~keV energy ranges after accounting for changes in attenuator states from A0 to A1 and back to A0 again. Horizontal lines marked A0 and A1 indicate the RHESSI attenuator state. The GOES 1--8~\AA~lightcurve is also plotted. The vertical {\it dashed} lines mark the start and end times of the EIS raster, while the vertical {\it solid} lines denote the times of the TRACE images in Figure~\ref{trace_hsi_eis_fov}. The first of these lines also corresponds to the time of the RHESSI image and spectrum (14:14:28--14:15:00~UT in the A0 state), and the time at which the EIS slit was rastering over the southeastern footpoint.}
\label{hsi_goes_ltc}
\end{center}
\end{figure}

The GOES C1.1 class flare under study occurred in NOAA AR 10978 on 2007 December 14 at 14:12 UT. Figure~\ref{trace_hsi_eis_fov} shows three images of the active region taken by the Transition Region and Coronal Explorer (TRACE; \citealt{hand99}) in the 171~\AA~passband at different stages during the flare. In the first panel, two bright EUV footpoints are visible in the northern end of the box which denotes the EIS field of view. The inset in the top left corner of the panel shows a close-up of the footpoints with overlaid contours of the 20--25~keV emission observed by RHESSI. After correcting for the 5~$\arcsec$ pointing offset between RHESSI and TRACE in both the solar X and solar Y directions, the two HXR sources align well with the EUV footpoints seen by TRACE. The second panel (taken at 14:25:26~UT) shows a small loop extending to the southeast from the left-hand footpoint, rather than connecting to the right-hand footpoint as one might expect. In the third panel (taken at 14:54:24~UT), a large overlying loop is observed to connect both footpoints with the opposite side of the active region. Figure~\ref{hsi_goes_ltc} shows the RHESSI and GOES lightcurves of the event with the times of each of the TRACE images denoted by solid vertical lines.

By January 2008 Hinode had lost the use of its X-band transmitter. As such, the amount of data being telemetered to the ground was severely reduced. Also, in November 2007, the detectors on RHESSI had been successfully annealed after 5 years of radiation damage, bringing them back to specifications comparable to early 2005. Therefore, December 2007 marks a unique timeframe when both Hinode and RHESSI datasets were near optimal. 

\begin{figure*}[!t]
\begin{center}
\includegraphics[height=16cm, angle=90]{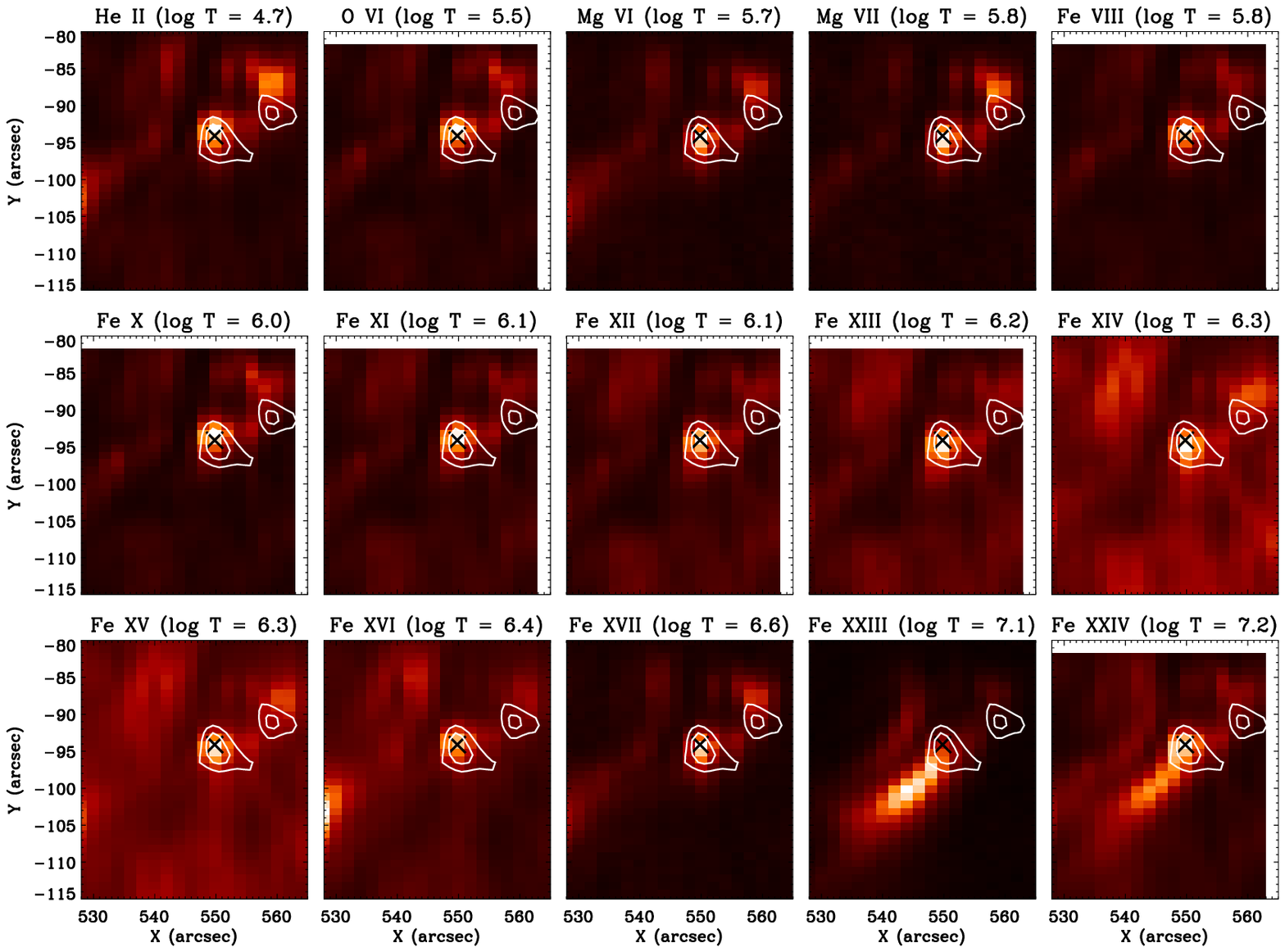}
\includegraphics[height=16cm, angle=90]{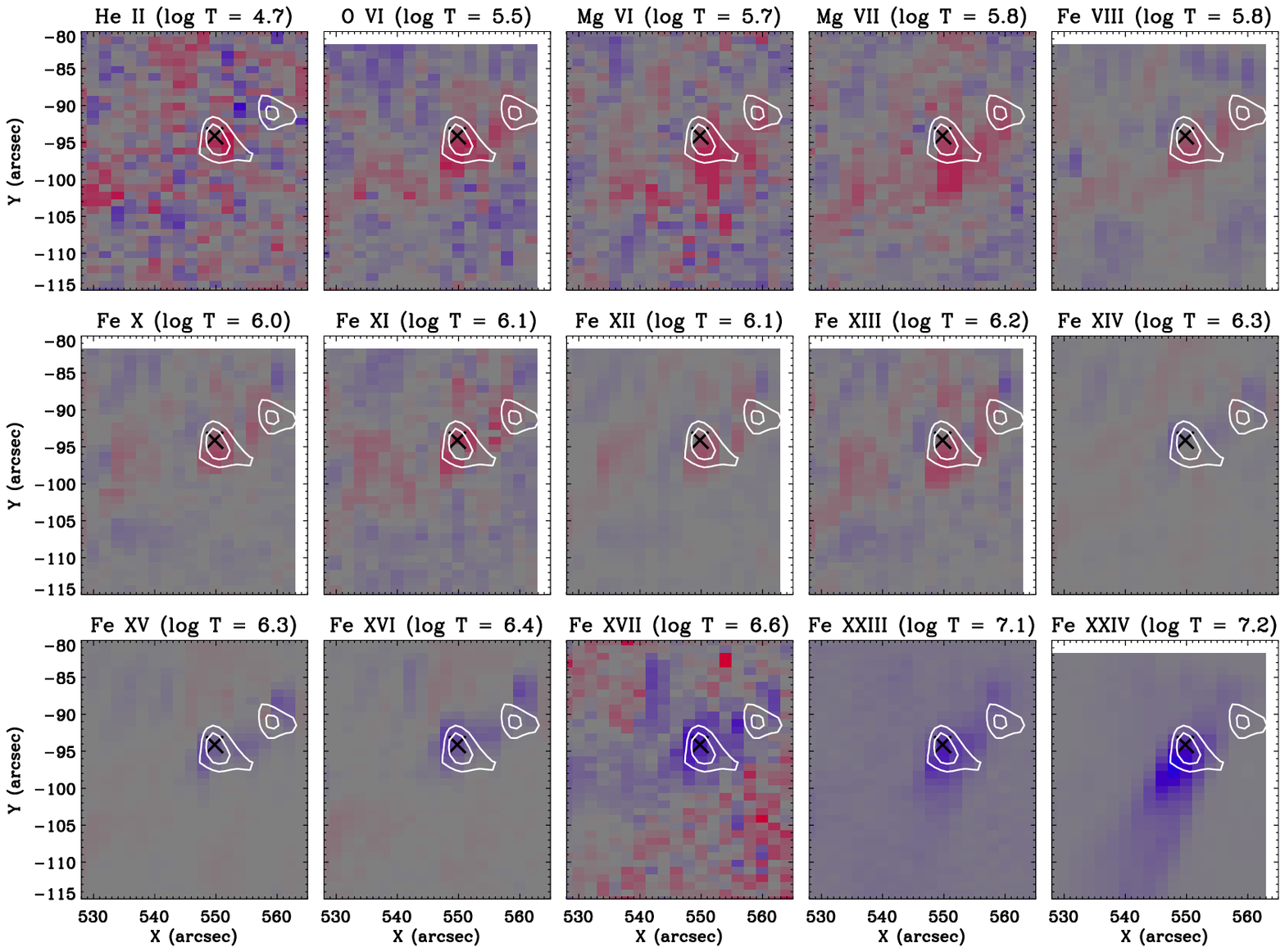}
\caption{{\it Top three rows}: Intensity maps in each of the 15 lines used in this study, ranging from 0.05--16~MK. Two footpoints are clearly visible, with the southeastern one being the brighter of the two. Overlaid are the 20--25~keV emission contours (at 60\% and 80\% of the maximum) as observed by RHESSI from 14:14:28--14:15:00~UT. The pixel marked with an `$\times$' within the HXR contour was the focus of a more detailed spectral analysis. {\it Bottom three rows}: Except for the \ion{Fe}{23} and \ion{Fe}{24} maps, the corresponding velocity maps for each of the above intensity maps. Red pixels denote material moving away from the observer, while blue pixels represent material moving towards the observer. The same RHESSI 20--25~keV contours are overlaid. All velocity maps are scaled to $\pm$150~km~s$^{-1}$. The \ion{Fe}{23} and \ion{Fe}{24} which are images formed over the enhanced blue wing of each line with the blue color scaled with the flux.}
\label{int_vel_maps}
\end{center}
\end{figure*}

\begin{figure*}[!ht]
\begin{center}
\includegraphics[height=18cm,angle=90]{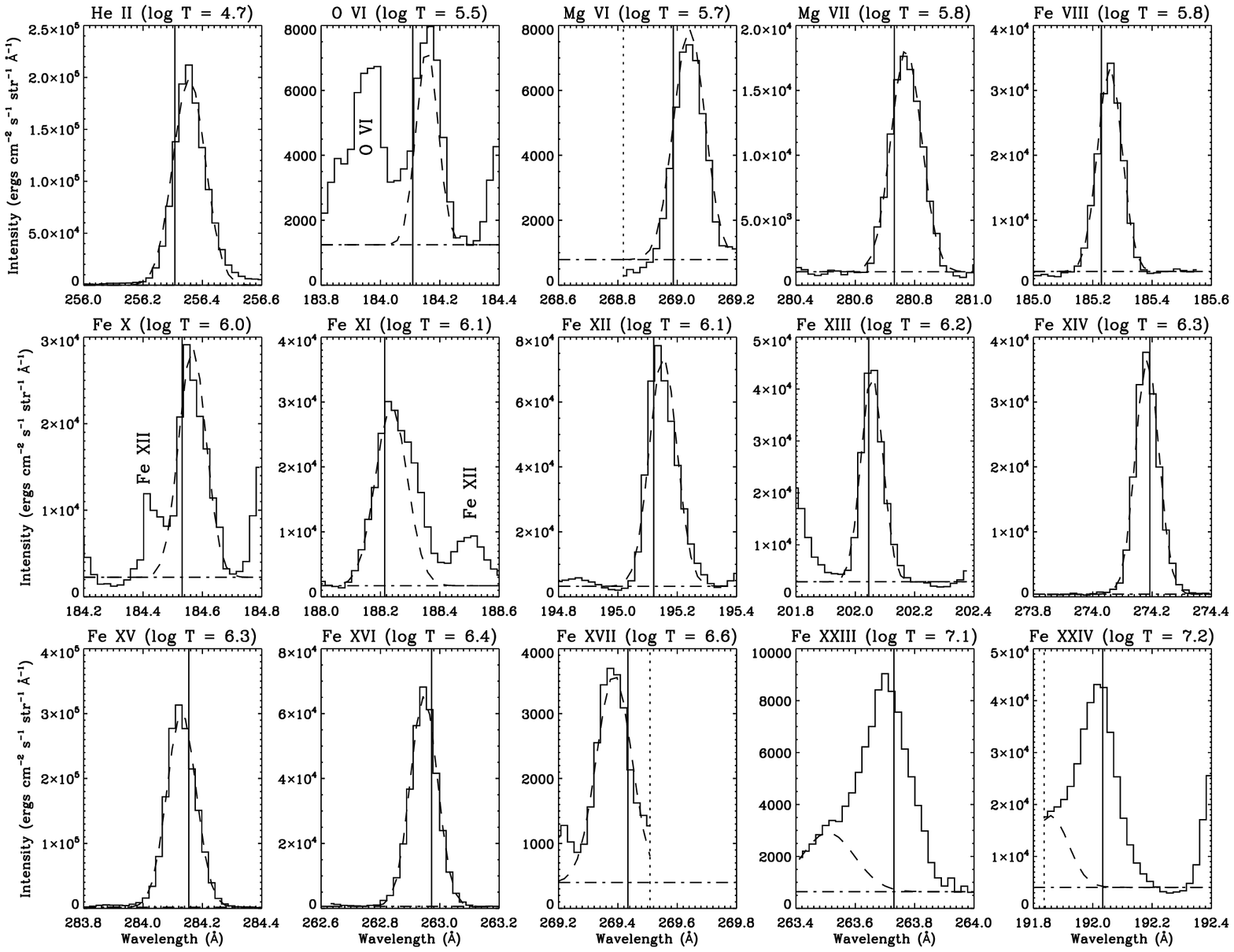}
\caption{Plots of each of the emission lines from the same spatial pixel from the southeastern footpoint taken at $\sim$14:14:51~UT. Only the Gaussian fits to the shifted components are shown and are denoted by {\it dashed} lines. The horizontal {\it dot--dashed} lines mark the background level. The vertical {\it solid} lines represent the rest wavelengths as measured from quiet-Sun regions, except in the cases of \ion{Fe}{23} and \ion{Fe}{24}. In these two panels, the vertical lines representing the rest wavelengths mark the centroid of the {\it dotted} Gaussian profiles obtained from integrating over the subsequent raster during the decay phase. In the \ion{Mg}{6}, \ion{Fe}{17}, and \ion{Fe}{24} panels the vertical {\it dotted} lines mark the edge of the spectral windows, beyond which no data were available.}
\label{eis_line_fit_fig}
\end{center}
\end{figure*}

\section{EIS Observations}
\label{eis_obs}
The observing study that EIS was running when the flare occurred (CAM\_ARTB\_RHESSI\_b\_2) was originally designed to search for active region and transition region brightenings in conjunction with RHESSI. Using the 2$\arcsec$ slit, EIS rastered across a region of the Sun from west to east covering an area of 40$\arcsec \times$143$\arcsec$, denoted by the rectangular boxes in Figure~\ref{trace_hsi_eis_fov}. Each slit position had an exposure time of 10~s resulting in an effective raster cadence of $\sim$3.5~minutes. These fast-raster studies are preferred for studying temporal variations of flare parameters while preserving the spatial information. Equally important though, is the large number of emission lines covering a broad temperature range. The observing study used 21 spectral windows, some of which contain several individual lines. The work presented here focuses on 15 lines spanning the temperature range 0.05--16~MK. Details of the lines, their wavelengths and peak formation temperatures are given in Table~\ref{line_data} (c.f. \citealt{youn07}), along with their measured Doppler velocities. The majority of these lines are well resolved and do not contain blends, thereby reducing ambiguities in the interpretation of their analysis. 

Intensity and velocity maps are shown in Figure~\ref{int_vel_maps} for the portion of the EIS raster containing the two footpoints in each of these 15 lines during the impulsive phase of the flare. Overlaid on each image are the contours of the 20--25~keV emission observed by RHESSI. Looking at the brighter southeastern footpoint in the first three rows of Figure~\ref{int_vel_maps}, there are no significant differences between images formed at temperatures lower than $\sim$4~MK. Images in the two hottest lines (\ion{Fe}{23} and \ion{Fe}{24}) however, show an overlying loop structure which had begun to fill with hot plasma. \\

\begin{table}[!b]
\begin{center}
\small
\caption{\textsc{\small{Ions, Wavelengths, Peak Formation Temperatures, and Measured Doppler Velocities of Emission Lines Used in This Work}}}
\label{line_data}
\begin{tabular}{lccc} 
\tableline
\tableline
\multicolumn{1}{c}{Ion}	&$\lambda$(\AA) &$T$ (MK)		&$v$ (km~s$^{-1}$)\\ 
\tableline
\ion{He}{2}	&256.32	&0.05	&21$\pm$12\\
\ion{O}{6}		&184.12	&0.3		&60$\pm$14\\
\ion{Mg}{6}	&268.99	&0.5		&51$\pm$15\\
\ion{Mg}{7}	&280.75	&0.6		&53$\pm$13\\
\ion{Fe}{8}	&185.21	&0.6		&33$\pm$17\\
\ion{Fe}{10}	&184.54	&1.0		&35$\pm$16\\
\ion{Fe}{11}	&188.23	&1.25	&43$\pm$15\\
\ion{Fe}{12}	&195.12	&1.25	&28$\pm$17\\
\ion{Fe}{13}	&202.04 	&1.5		&39$\pm$14\\
\ion{Fe}{14}	&274.20	&2.0		&-22$\pm$12\\
\ion{Fe}{15}	&284.16	&2.0		&-32$\pm$8\\
\ion{Fe}{16}	&262.98	&2.5		&-39$\pm$20\\
\ion{Fe}{17}	&269.17	&4.0		&-69$\pm$18\\
\ion{Fe}{23}	&263.76	&12.5	&$<$-252$\pm$32\\
\ion{Fe}{24}	&192.03	&16.0	&$<$-268$\pm$28\\
\tableline
\normalsize
\end{tabular}
\end{center}
\end{table}

\begin{figure}[!t]
\begin{center}
\includegraphics[height=9cm,angle=90]{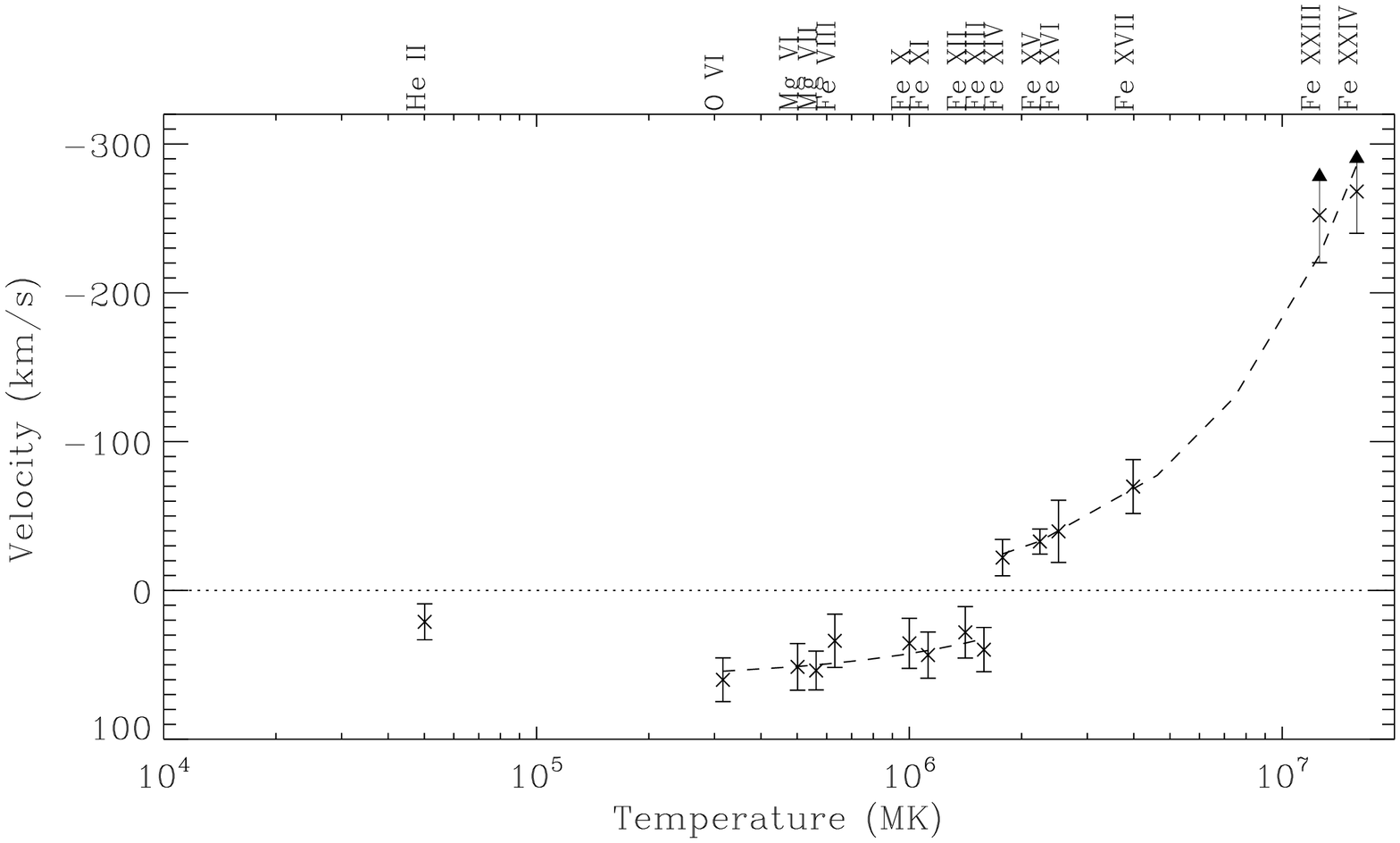}
\caption{Plasma velocity from a flare footpoint at $\sim$14:14:51~UT as a function of temperature for each of the emission lines used in this study. The {\it dashed} lines represent a weighted least squares fit to the data points from 0.5 to 1.5 MK and 2.0 to 16 MK.}
\label{vel_temp_fig}
\end{center}
\end{figure}

\subsection{Emission Line Fitting and Velocity Analysis}
\label{line_fit_vel_anal}

To determine the magnitude, velocity, and spatial distribution of the evaporating plasma, a measurement of a reliable rest wavelength is critical for each line under study. Following from previous work \citep{mill08}, the rest wavelength of each emission line of interest was established from the corresponding spectrum summed over the bottom half of the EIS raster shown in Figure~\ref{trace_hsi_eis_fov}. This area encompassed both quiet-Sun and quiescent active region emission.

In the case of the two hottest lines (\ion{Fe}{23} and \ion{Fe}{24}), little or no quiescent emission is detected. Instead, a weak \ion{Fe}{11} line is detected close to the expected rest value of \ion{Fe}{24} \citep{youn07} and \ion{Fe}{23} yields to \ion{Ar}{15} \citep{delz08}. Therefore an alternate method must be used when determining a rest value against which to measure Doppler velocities for these two lines. The commonly accepted practice for such high-temperature lines is to determine their centroid positions later in the flare after flows have ceased. For EIS data, however, this introduces other problems. As the EIS detectors are sensitive to changes in temperature during Hinode's orbit, this orbital variation needs to be accounted for before comparing measurements from different stages of the spacecraft orbit. In the case presented here the raster durations are short when compared to the orbital period of 96~minutes ($\sim$3\%). Therefore, to establish rest wavelengths for \ion{Fe}{23} and \ion{Fe}{24}, the raster succeeding the one during the impulsive phase was used. As this subsequent raster was taken during the beginning of the decay phase, the assumptions can be made that the spectra are still dominated by the hottest plasma (rather than the cooler, blended lines), and the plasma is predominantly at rest as the bulk of the energy deposition had ceased. Using this subsequent raster 3.5 minutes later also assures that orbital variation effects are minimized. 

The spectrum in each emission line window was analyzed as follows to determine the Doppler velocities of the shifted compnents: The centroid of each line in each pixel within the raster from the impulsive phase was calculated by fitting a Gaussian profile plus constant background to the data. Doppler velocities for each line were then measured relative to their respective rest wavelengths. Uncertainties in velocity were determined from the 1$\sigma$ widths of the rest and shifted components added in quadrature. This is an overestimate of the uncertainty since the Gaussian peaks can be located more accurately. Corrections were performed for variations in the centroid position along the EIS slit. 

The bottom three rows in Figure~\ref{int_vel_maps} show the corresponding velocity maps created using the above method with contours of the HXR emission observed by RHESSI overlaid on each panel. For ions \ion{He}{2}--\ion{Fe}{8} (0.05--1.5~MK), footpoint emission is clearly redshifted, while material at \ion{Fe}{14-XXIV} temperatures (2--16~MK) appears blueshifted. The \ion{Fe}{23} and \ion{Fe}{24} velocity maps were formed using a different method, as true velocity maps become distorted by blended lines in regions where cooler emission begins to dominate. In these cases, images were formed over the enhanced blue wing of each of the two lines. It is not possible to determine the magnitude of the flows using this method, but it clearly shows that the spatial distribution of upflowing material is cospatial with the HXR emission, and is unaffected by blended lines.

Figure~\ref{eis_line_fit_fig} shows the 15 emission line profiles taken from the same spatial pixel within the HXR footpoint in each of the panels in Figure~\ref{int_vel_maps} (marked with an `$\times$'). The time at which the slit of the EIS spectrometer was rastering over this pixel was 14:14:51$\pm$20~UT. All the lines formed at temperatures less than 4~MK show a symmetric Gaussian profile completely shifted with respect to their rest wavelengths. In some panels, other Gaussian components from neighboring lines in the passband can also be seen, although, for clarity, only fits to the lines of interest are shown. The two highest temperature line profiles, \ion{Fe}{23} and \ion{Fe}{24}, in contrast, show large blue-wing asymmetries and were therefore each fitted with two Gaussian components. The dominant components were each found to have centroid positions close to their rest wavelengths indicative of predominantly stationary material. The Gaussian component fitted to the blue-wing was indicative of weaker, high-velocity upflows. In each case, the spectral windows were not large enough to accommodate such large blueshifts of the lines, so the derived Doppler velocities of $-$252$\pm$32 and $-$268$\pm$28~km~s$^{-1}$ should be considered as lower limits on the absolute values. It is also worth noting that the EIS pixel that is the focus of this work is not unique; many of the neighboring pixels within the footpoint show similar profiles with a dominant stationary component in the two hottest lines.

\begin{figure}[!t]
\begin{center}
\includegraphics[height=8.5cm,angle=90]{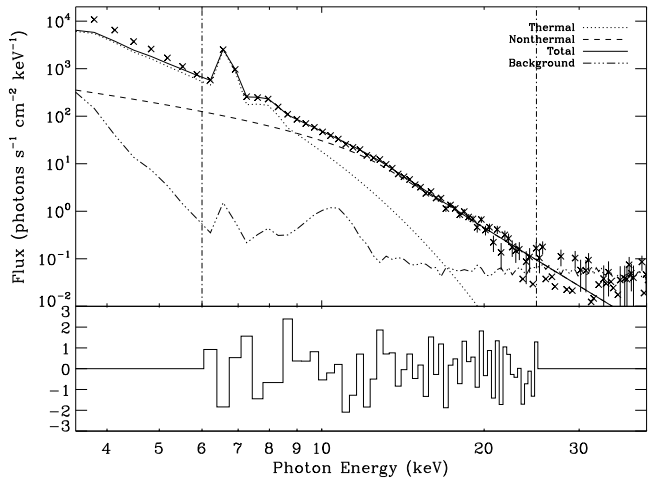}
\caption{RHESSI photon spectrum from detector 4 taken during the time that blueshifted emission was observed by EIS (14:14:28--14:15:00~UT). The {\it dotted} line represents the best fit to the thermal component while the {\it dashed} line represents the thick-target component. The {\it solid} line shows the sum of the two components and the {\it triple-dot-dashed} line marks the background. The two vertical {\it dot-dashed} lines mark the energy range over which the spectral model was fitted to the data. Beneath the spectrum are the associated residuals from the least-squares fit normalized to 1$\sigma$ at each energy.}
\label{hsi_spec_fig}
\end{center}
\end{figure}

Figure~\ref{vel_temp_fig} shows the derived line-of-sight velocities from this pixel as a function of the peak formation temperature for each of the lines listed in Table~\ref{line_data}. Assuming a linear relationship between velocity ($v_{up}$ and $v_{down}$) and temperature ($T$) of the form $v=A+BT$, where $A$ and $B$ are constants, a least squares fit was applied to the both blueshifted and redshifted data points and their associated uncertainties (excluding \ion{He}{2}). (Note that Figure~\ref{vel_temp_fig} is on a log-normal scale.) All emission from \ion{O}{6} to \ion{Fe}{13} (0.6--1.5~MK) was redshifted by $v_{down}$ (km~s$^{-1}$) $\approx$ 60$-$17 $T$ (MK). The \ion{He}{2} line was found to be redshifted by 21$\pm$12~km~s$^{-1}$ which is slightly lower than the other redshifted values. This is possibly due to the fact that \ion{He}{2} is optically thick and hence the line-of-sight velocities are underestimated. These redshifts are interpreted as evidence for the downward-moving `plug' generated by the overpressure of the rising material above. The evaporated material itself is only observed in emission lines formed at temperatures higher than 2~MK (\ion{Fe}{14}). Upflow velocity was found to scale as $v_{up}$ (km~s$^{-1}$) $\approx$ 8$-$18 $T$ (MK). The direction of the derived flow velocities also changes sharply between \ion{Fe}{13} and \ion{Fe}{14}. This corresponds to a temperature range of $\sim$0.5~MK, which is narrower than the width of the associated contribution functions of either of these two lines. 

\section{RHESSI Imaging and Spectral Analysis}
\label{rhessi}

\begin{figure}[!t]
\begin{center}
\includegraphics[width=8.5cm]{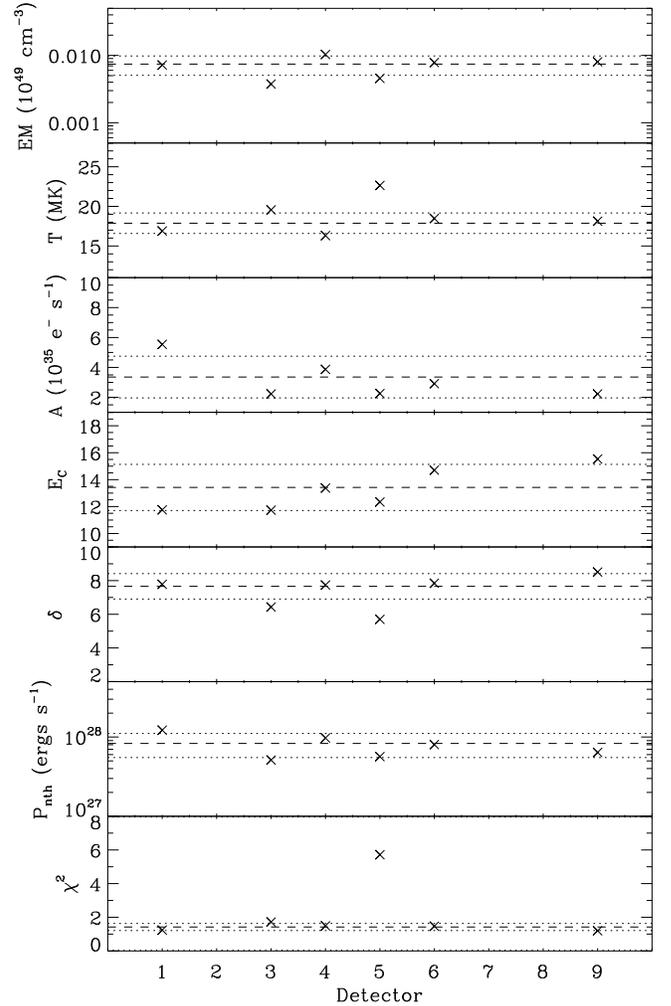}
\caption{Plot of the spectral fit parameters for each RHESSI detector. From top to bottom: emission measure ($EM$), temperature ($T$), total integrated electron flux ($I$), low-energy cutoff ($E_{c}$), spectral index ($\delta$), power in nonthermal electrons ($P_{nth}$), and reduced $\chi^{2}$. The horizontal {\it dashed} and {\it dotted} lines on each panel mark the mean and 1$\sigma$ values of each parameter, respectively.}
\label{hsi_sep_det}
\end{center}
\end{figure}

Spatially-integrated RHESSI HXR spectra were used to determine the flux of electrons (in ergs~s$^{-1}$) thought to be responsible for the evaporation seen with EIS, and the RHESSI images determine the footpoint area to give the energy flux in ergs~cm$^{-2}$~s$^{-1}$. RHESSI spectra were formed over the time range 14:14:28--14:15:00~UT for detectors 1, 3, 4, 5, 6, and 9 individually. This time interval corresponds to when the EIS slit rastered over the southeastern footpoint and before RHESSI's thin attenuators came into place over the detectors (see Figure~\ref{hsi_goes_ltc}). Detectors 2 and 7 are known to be poorly sensitive to emission below 20~keV and were therefore not included. During this phase of the event, data from detector 8 were also not used due to interference while RHESSI's antenna was transmitting. Compiling spectra for each detector individually allows the most up-to-date corrections for pulse pileup and gain offset available in the OSPEX software package to be used. The data were fitted with the sum of an isothermal component that dominated at low energies and a thick-target model dominating at higher energies. Figure~\ref{hsi_spec_fig} shows the photon spectrum for detector 4 with the associated fits and residuals while Figure~\ref{hsi_sep_det} shows the fit parameters for each of the six detectors used. Detector 5 was found to give consistently higher values for $\chi^{2}$ as the calibration does not appear to be well known at this time, and was therefore omitted from any further calculations. Taking the mean and standard deviation of each fit parameter for each component across the 5 individual detectors currently provides the best estimate of the parameter and its uncertainties. The isothermal fits yielded a temperature ($T$) and emission measure ($EM$) of 17$\pm$1~MK and 7$\pm$2$\times$10$^{46}$~cm$^{-3}$, respectively. The low-energy cutoff to the assumed power-law electron spectrum was found to be, $E_{c} \leq$ 13$\pm$2~keV with a spectral index, $\delta$, of 7.6$\pm$0.7. 

The combination of high-resolution images and spectra from RHESSI allows a measurement of the flux of nonthermal electrons responsible for driving chromospheric evaporation to be determined. Given the values of the low-energy cutoff and the spectral index of the electron distribution, the total power contained in the electron beam can be calculated using:

\begin{equation}
P_{nth}(E \geq E_c) = \int_{E_c}^{\infty} E F(E) dE~~{\mbox{ergs~s$^{-1}$}},
\end{equation}
where $F(E)$ = $CE^{-\delta}$~electrons~s$^{-1}$~keV$^{-1}$, and $C$ is a normalization constant proportional to the total integrated electron flux, $I$. Using the above values $P_{nth} \gtrsim$ 8$\pm$3$\times$10$^{27}$~ergs~s$^{-1}$. 

In order to compare observations with the predictions of theory the total energy flux (in ergs~cm$^{-2}$~s$^{-1}$) of nonthermal electrons must be established. This requires knowledge of the footpoint areas that can be derived frrom RHESSI images. Knowing that the highest energy emission was predominantly nonthermal, HXR images were formed over the same time range as the spectrum using the CLEAN algorithm, from 20--25~keV and using detectors 1--6. Two HXR sources were identified which aligned with the footpoint emission detected by TRACE (shown in the inset of the first panel of Figure~\ref{trace_hsi_eis_fov}) and EIS (Figure~\ref{int_vel_maps}). As an approximation, summing over all pixels within the 60\% contour of the 20--25~keV CLEAN images, yielded an area of $\sim$3$\times$10$^{17}$~cm$^{-2}$ for the sum of both footpoints. A similar value was found by summing over all the pixels within the 40\% contour of the TRACE image. Both of these percentage levels were chosen as they comfortably distinguish between footpoint and background emission. It is known that CLEAN can overestimate source areas by as much as a factor of 10 compared to other image reconstruction algorithms \citep{schm07}, thereby placing a lower limit on the value of the electron flux, assuming a filling factor of unity. \cite{denn09} have compiled a detailed comparison of how each available algorithm can be optimized to provide reliable estimates of source sizes. Using the CLEAN algorithm, for example, the authors compute the moments for individual CLEAN components as functions of the azimuthal angle about the source. The moments then define the parameters of the equivalent elliptical Gaussian, and can be used to determine the source area within 1$\sigma$ of the centroid. Applying this technique to the HXR images for this event resulted in a combined footpoint area of 1$\times$10$^{17}$~cm$^{-2}$; a factor of three smaller than the above approximation. A similar value was found using the Pixon algorithm. Dividing $P_{nth}$ by this footpoint area gives a flux value of $F_{nth} \gtrsim5 \times$10$^{10}$ ergs~cm$^{-2}$~s$^{-1}$, which is comfortably above the limit that \cite{fish85c} stated is needed to drive explosive chromospheric evaporation. \\

\section{DISCUSSION AND CONCLUSIONS}
\label{conc}

This is the first detailed study of chromospheric evaporation carried out using data from Hinode/EIS during the impulsive phase of a C-class flare. Previous studies (\citealt{anto83,zarr88,canf87,acto82,bros04,teri03,mill06a,mill06b}) were limited by the detection of blueshifted emission in a single high temperature line, and often without any spatial information or knowledge of the driving electron beam. Coordinated observations between EIS and RHESSI are now able to overcome these limitations with RHESSI providing a measure of the electron beam energetics. The observations presented here indicate a clear case of explosive chromspheric evaporation. High-temperature upflows and lower-temperature downflows measured by EIS were found to be both spatially and temporally correlated with HXR emission observed by RHESSI. 

The key findings of this study can be summarized as follows: (1) All emission lines within a footpoint pixel exhibit a complete symmetric shift of their respective line profiles with the exception of \ion{Fe}{23} and \ion{Fe}{24}. These two hottest lines each show a dominant stationary component with a large blue-wing enhancement indicative of upflow velocities $>$200~km~s$^{-1}$. (2) Upflow velocity was found to be strongly dependent on temperature ($v_{up}$ km~s$^{-1} \approx $8$-$18 $T$ MK between 2--16~MK). (3) The velocity of the downward-moving plug also shows a linear dependence upon temperatures ($v_{down}$ km~s$^{-1} \approx $60$-$17$T$ MK between 0.5--1.5~MK). (4) Downflows are detected at much higher temperatures (up to 1.5~MK) than previously observed or predicted by any current chromospheric evaporation model (with the exception of \citealt{liu08}). (5) The division between the temperature of upflowing and downflowing material occurs over a very narrow range ($<$0.5~MK).

The most significant finding presented here is the discovery of dominant stationary, high-temperature emission (\ion{Fe}{23} and \ion{Fe}{24}; $>$12~MK) at a footpoint during the impulsive phase. The stationary component of the \ion{Ca}{19} line from spatially integrated BCS spectra in previous works was believed to be from the top of the flare loop where evaporated material had collected, or was moving perpendicular to the line of sight \citep{dosc05}. The fact that this component dominated the line profile from flare onset contradicted the basic evaporation model which predicts that the first detection of flaring plasma should be completely blueshifted \citep{li89}. \cite{dosc05} stated the possibility that current instrumentation is not sensitive enough to detect the earliest blueshifted emission, and by the time emission levels had risen sufficiently, the flare loops had already been filled. Even though the EIS slit was not rastering over the footpoint at the flare onset in this case, the presence of any stationary, high-temperature plasma at a footpoint during the impulsive phase indicates a severe discrepancy with the standard flare model. Any chromospheric material that is subjected to heating during a flare should, in principle, rise or expand due to the pressure gradient along the overlying loop. It is therefore unclear at present which mechanism could be responsible for keeping the bulk of the hottest material at rest. The possibility must be acknowledged, however, that a small, unresolved loop or magnetic structure exists within the spatial pixel in question that could be responsible for confining the hottest emission.

The detection of temperature dependent upflows was previously reported by \cite{imad07} who also used EIS data. Emission lines formed below 1~MK showed a weak velocity dependence on temperature (values $<$50~km~s$^{-1}$), while higher temperature lines exhibited a stronger dependence up to 150~km~s$^{-1}$ in \ion{Fe}{15}. Although these measurements were made during the decay phase of an X-class flare, spatially they were detected in a weak plage region. In their paper, the authors do not state how these flows relate to the flare, what their driving mechanism was, or why the temperature dependency exists. In the work presented here, the observed flows are clearly both spatially and temporally correlated to HXR emitting regions and can be quantifiably linked through RHESSI observations to an electron beam during the impulsive phase. The expected dependence of velocity on temperature of evaporating material has been predicted by \cite{fish85c}. Under the assumption that pre-flare conditions were in static equilibrium, chromospheric material heated to different temperatures will be subject to different pressure gradients relative to the overlying corona and therefore rise at different rates. 

The downward-moving plug is more difficult to interpret. In this case, redshifts are detected in emission lines formed at much higher temperatures than previously expected or reported (up to 1.5~MK). \cite{mill06a} detected this plug at \ion{He}{1} and \ion{O}{5} temperatures (0.03 and 0.25~MK, respectively) during the impulsive phase of an M-class flare using CDS data in agreement with the model of \cite{fish85a}. Redshifted \ion{Fe}{15} (2~MK) emission was also detected at a footpoint during a B-class flare by \cite{mill08} using EIS data. In this instance the nature of such high-temperature downflows was unclear. Coordinated RHESSI observations implied that there was no detectable nonthermal component to the X-ray spectrum indicative of the electron beam required to generate the localized pressure enhancement in the chromosphere. The assumption was therefore made that a syphon flow may have been in place. The similarity between these flows and the ones presented here suggest that such high-temperature downflows may be a common characteristic of the impulsive phase of solar flares.

There are two possible explanations for material up to 1.5~MK to be redshifted as the electrons deposit their energy. Either the electrons lose their energy in the upper transition region rather than the chromosphere, or the chromosphere was being heated as it recoils. Although the soft spectral index measured by RHESSI ($\delta=7.6$) might imply that the bulk of the electrons would be stopped at higher altitudes, it is unlikely that the preflare densities at this height would be high enough. Following from \cite{fish85c}, \cite{liu08} has developed a model of the chromospheric response to explosive evaporation that incorporated continuous electron deposition. This new model is now able to explain both the high-temperature downflows and the rate of recoil. The earlier \cite{fish85c} model only considered the chromospheric response to a single burst of energetic electrons, whereas \cite{liu08} showed that as energy deposition continues throughout the impulsive phase, the underlying chromosphere is also heated to temperatures approaching 2~MK, in agreement with the observations presented here. The temperature at which the sign of the Doppler velocity changes between positive and negative may therefore not necessarily be an indicator of the depth at which the bulk of the electron energy is deposited but may rather depend on the duration and/or magnitude of the energy deposition at the footpoints.

\section{Future Work}
\label{future}

This paper has focused primarily on measuring Doppler shifts of EUV emission lines during the impulsive phase of this event. However, thanks to the high raster cadence of the observing study and the extensive list of emission lines, line intensities, shifts and widths can all be studied as a functions of time and temperature throughout the event. The multitude of emission lines could also allow for a more accurate evaluation of momentum balance between the evaporating and condensing material. This study also contains several pairs of density-sensitive line ratios which could be used to investigate density enhancements and variations throughout the flare. Therefore, operating EIS in `fast-raster' mode with a moderate field of view and a carefully chosen list of emission lines not only increases the probability of successfully observing a flare, but also enables the study of a wide range of fundamental flare physics. Coincident RHESSI observations also provide information on the location and magnitude of both the thermal and nonthermal X-ray emission at high spatial, spectral and temporal resolution. Continued observations between EIS and RHESSI therefore allow for a more comprehensive analysis of the dynamics and energetics of solar flares during the rise of Cycle 24.

\acknowledgments
This research was supported by an appointment to the NASA Postdoctoral Program at the Goddard Space Flight Center, administered by Oak Ridge Associated Universities through a contract with NASA. Hinode is a Japanese mission developed and launched by ISAS/JAXA, collaborating with NAOJ as a domestic partner, NASA and STFC (UK) as international partners. Scientific operation of the Hinode mission is conducted by the Hinode science team organized at ISAS/JAXA. Support for the post-launch operation is provided by JAXA and NAOJ (Japan), STFC (U.K.), NASA, ESA, and NSC (Norway). The authors would like to thank Len Culhane, Dave Williams, C. Alex Young, and Claire Raftery for their very helpful and insightful discussions.

\bibliographystyle{apj}
\bibliography{ms}


\end{document}